\documentclass[10pt]{iopart}
\usepackage{graphicx}
\begin{document}

\title[Direct measurement of correlations in twin photon beams]{Simple direct
measurement of nonclassical joint signal-idler photon-number
statistics and correlation area of twin photon beams}

\author{Ond\v{r}ej Haderka\footnote[3]{To whom correspondence should be addressed
(haderka@sloup.upol.cz)}, Jan Pe\v{r}ina, Jr., and Martin Hamar
 }

\address{Joint Laboratory of Optics of Palack\'{y} University and Institute of
Physics of Academy of Sciences of the Czech Republic,
Av. 17. listopadu 50A, 772 00 Olomouc, Czech Republic}

\begin{abstract}
The measurement of joint signal-idler photon-number distribution
of a field obtained from spontaneous parametric downconversion
using an intensified CCD camera is presented. It is shown that a
classicality criterion is violated directly by the measured data.
Characteristic dimensions of the area of correlation are
determined in the same experimental setup.
\end{abstract}


\submitto{\JOB}

\maketitle

\section{Introduction}
In recent years an enormous number of applications of correlated
photon pairs obtained by the process of spontaneous parametric
down-conversion has emerged. Such photon pairs are quantum
correlated in various quantities including photon number,
position, momentum, energy, polarization, angular orbital momentum
etc. These correlations are responsible for highly nonclassical
properties of the generated light fields that enabled to
successfully test the fundamental rules of quantum theory (see,
e.g., \cite{Zeilinger99} for a review) as well as to co-found a
whole new field of quantum communications. Quantum key
distribution (see, e.g., \cite{Gisin02} for a review),
teleportation \cite{Bennett93}, dense coding \cite{Mattle96} or
entanglement-swapping \cite{Bennett93,Zukowski93}, among others,
are now well established methods that make use of nonclassical
correlations between the signal and idler photons.

While the process of spontaneous parametric down-conversion (SPDC)
itself has been thoroughly studied for many years
\cite{Milburn95,Mandel95,Perina94}, still new approaches to the
characterization of the correlations are being developed. The
first experiments to obtain correlated photon pairs used bulk
nonlinear crystals pumped by cw ion lasers \cite{Hong87} (omitting
pioneering approach based on resonant fluorescence
\cite{Aspect81}). Since efficiency of the process is quite low,
the mean number of pairs per detection interval has always been
much lower than one. Despite the fact that the real photon-number
distribution of photon pairs is Poissonian (Bose-Einstein) in
multi-mode case (single-mode case), the output state could be
described as a superposition of vacuum and one-pair state due to a
low mean photon number. The idea of using SPDC for a probabilistic
source of single-photon states \cite{Hong86} emerged from this
form of the output state. Availability of ultrafast tunable pulsed
lasers was welcomed by the researchers as it enabled to produce
down-converted pairs synchronized in time and also with the
detection equipment, as well as to achieve considerably higher
mean photon numbers in a time window given by a pump-pulse
duration time. Also development of new materials featuring several
orders of magnitude higher nonlinear efficiencies
\cite{Tanzilli02} in comparison with bulk crystals contributed in
the same direction. Nowadays even SPDC sources pumped by laser
diodes are available \cite{Alibart04}.

On the other hand, the analysis of photon-number statistics of the
down-converted fields is not a simple task because traditional
detectors offer only single-photon sensitivity ---
photomultipliers or avalanche photodiodes do not usually yield an
information about the number of photons detected in a detection
interval owing to a large noise introduced in the signal
amplification process. They work just as trigger detectors
announcing an impingement of some light quanta. Recently there has
been devoted a great deal of interest to overcome this limitation.
Two main classes of approaches in construction of photon-number
resolving devices have been developed.

The first class relies on getting the amplification process under
better control. This is possible using special detection
structures cooled down to low temperatures. Two approaches seem
currently the most promising. A visible light photon-counter
(VLPC) \cite{Kim99} can resolve small photon numbers in the
visible light range with a very high quantum efficiency at
temperatures of 6-7K but it shows a quite high dark-count noise. A
super-conducting transition-edge sensor micro-calorimeter
\cite{Miller03} on the other hand has a low noise and covers a
wider spectral range but its quantum efficiency is currently
lower. It also requires sub-1K temperatures. While development of
these devices undergoes a rapid progress they are still quite
demanding in operation.

The second class of devices is based on a beam division and it
makes use of the fact that photons in the state to be analyzed
behave independently on a beam-splitter. When the number of paths
offered to a photon is much larger than the number of all photons,
each photon takes a different route with a high probability and
can be detected alone with a trigger detector. The total number of
trigger detections then yields an information about the number of
photons \cite{Paul97}. While many-detector devices are impractical,
several configurations using only one or two trigger detectors
have been devised and constructed
\cite{Banaszek03,Achilles03,Achilles04,Fitch03,Haderka04,Rehacek03} utilizing
fiber delay-loops to translate different spatial paths into a
time-multiplexed signal. Still, the number of photons to be
analyzed by these devices is limited by the available number of
time-multiplexed channels to rather low photon numbers. In our
previous publication \cite{Haderka05} we have used a different device
for the measurement of photon-number statistics of photon pairs
--- an intensified CCD camera (iCCD). iCCD can be seen as a
massively multichannel device where each pixel serves as a trigger
detector with single-photon sensitivity. Since the total detection
efficiency in our previous experiment was quite low, we have used
reconstruction methods based on an expectation-maximization
algorithm to get the field emerging from the process of SPDC and
to show its nonclassical character (a new non-classicality
criterion has been introduced).

In this paper we show that with an improved detection efficiency
the classicality criterion is violated even directly by the
measured data. In addition, we employ the spatial resolution
provided by the iCCD to measure directly the area of correlation
of the twin photon beams. We note that an iCCD camera has been
used for the measurement of spatial correlations in SPDC in a
different experimental configuration previously \cite{Jost98} and
scientific CCD cameras have been applied for the analysis of
strong down-converted fields in \cite{Jedrkiewicz04,Lantz04}.
Measurement of the photon-number statistics in a summed
signal-idler collinear beam has also been accomplished using a
VLPC detector \cite{Waks04}.

\section{Experimental setup}

To measure the joint signal-idler photon-number distribution, we
use a simple experimental setup. Photon pairs are generated by a
type-I SPDC process in a 5~mm long LiIO$_3$ bulk nonlinear
crystal. Using 400~nm pumping, wavelength-degenerate pairs are
generated at a cone-layer with a vertex angle of 31~deg behind the
crystal (see Figure~1).

\begin{figure}
\begin{center}
\includegraphics[width=0.5\textwidth]{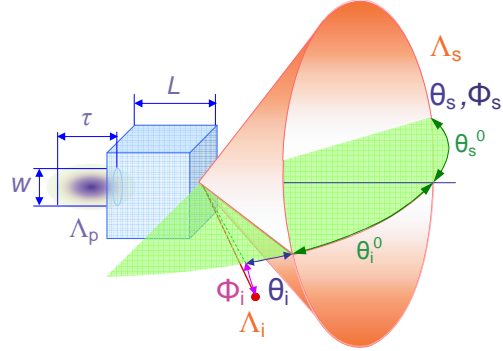}
\end{center}
\caption{\label{crystal}Degenerate ($\Lambda_s=\Lambda_i$) photon
pairs occur at opposite points of a cone layer behind the crystal.
There is a certain spread in the position (given by
$\Delta\Theta_i,\Delta\Phi_i$) occupied by an idler photon being
correlated with a signal photon detected at $\Theta_s, \Phi_s$.
This spread depends on a number of quantities including pump pulse
width, duration, spectrum, beam shape and crystal properties.}
\end{figure}

In our experimental arrangement, we let one part of the cone layer
to impinge directly on the photocathode of the iCCD camera while
the opposite part of the cone layer is reflected on a mirror
placed very close to the crystal to minimize path difference. The
width of the cone layer is determined by filtering both beams
using a 20~nm (FWHM) wide interference filter, thus accepting
slightly nondegenerate pairs as well. Two additional edge filters
(high-pass above 750~nm) are used to remove the majority of
laboratory stray light, scattered pump beam and fluorescence from
the crystal.

\begin{figure}
\begin{center}
\includegraphics[width=0.8\textwidth]{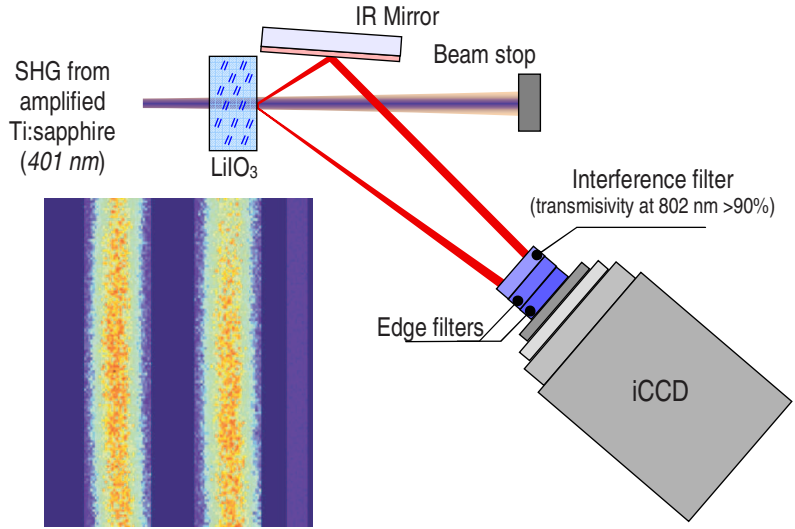}
\end{center}
\caption{\label{scheme}Scheme of the experiment. The iCCD camera
is placed further from the crystal in reality than in the scheme.
The inset shows an accumulated image from the iCCD camera after
the exposure of 240,000 frames.}
\end{figure}

The pump beam is obtained as second-harmonic of a femtosecond
train of pulses from an amplified Ti:sapphire laser (Coherent
Mira/RegA). The pulses are 200~fs long and their repetition rate
is controlled by the amplifier and set to 11~kHz in our case.
Their second-harmonic is produced in a 2-mm long BBO crystal and
pulses of energy up to 1.3~$\mu$J are obtained. This is sufficient
to generate several thousands of pairs to all spatial modes
accepted by the interference filter; we therefore attenuate the
pumping beam using a variable attenuator.

In the software of the camera, three regions of interest are
defined, two for the signal and idler strips and a third one that
serves for monitoring the noise level (see inset in
Figure~\ref{scheme}). Frames of the camera are transferred
one-by-one to a PC that performs image processing based on double
thresholding and centroid-finding and counts photon detections in
all three detection strips. The maximum overall quantum efficiency
of the iCCD camera has been found to be about 14\%. Total
detection efficiency including filters, mirror, and crystal output
face has been estimated to reach approximately 7\%.

\section{A direct measurement of the nonclassical character of
joint signal-idler photon-number distribution}

In our previous publication \cite{Haderka05} we have shown that even
though the marginal signal (idler) distributions are almost
Poissonian, as it should be for a multimode SPDC, the joint
signal-idler photon-number distribution $ f(c_S,c_I) $ shows
correlations in photon numbers. These correlations have been
observed (i) by comparing the measured $ f(c_S,c_I) $ with that
one obtained as a direct product of the marginal distributions (as
if they were independent) and (ii) by computing the correlation
coefficient $ C_p $,
\begin{equation}
  C_p = \frac{ \langle \Delta n_S \Delta n_I \rangle }{
  \sqrt{ \langle (\Delta n_S)^2
    \rangle \langle (\Delta n_I)^2 \rangle }} ,
        \Delta n_i = n_i - \langle n_i \rangle,
      i=S,I,
      \label{corr}
\end{equation}
that reached the value of $0.0435\pm0.008$. We have also
introduced the following inequality:
\begin{equation}
 p(n_S,n_I) \le \frac{n_S^{n_S}}{n_S!} \exp(-n_S) \, \frac{n_I^{n_I}}{n_I!}
  \exp(-n_I),
 \label{crit}
\end{equation}
that must be fulfilled by any photon-number distribution $
p(n_S,n_I) $ originating in a classical field
\cite{Hillery1987,Hillery1985}. This classicality criterion can be
readily obtained starting from a photodetection equation
\cite{Perina1991} for $ p(n_S,n_I) $ and assuming a nonnegative
distribution function of integrated intensities. Mainly due to a
low detection efficiency we were not able to show a violation of
this inequality by the measured photon-number distribution $
f(c_S,c_I) $.  We have shown, however, that the reconstructed
photon-number distribution $ p(n_S,n_I) $ at the output plane of
the crystal shows the maximum violation of the criterion
(\ref{crit}) by 49.2 standard deviations.

\begin{figure}
\begin{center}
\includegraphics[width=\textwidth]{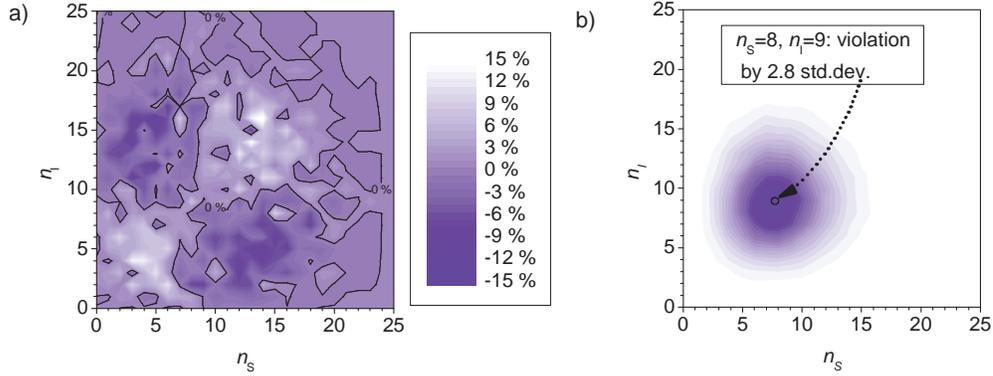}
\end{center}
\caption{\label{violation}(a) Difference between the measured
joint signal-idler photon-number distribution  $ f(n_S,n_I) $ and the
calculated joint signal-idler photon-number distribution given by a
direct product of the signal and
idler marginal distributions. Solid
line denotes zero contour. (b) Color-density graph of the
measured joint signal-idler photon-number distribution. Contour line in the graph
surrounds the points violating the classical inequality given
  in Eq.~\ref{crit}.}
\end{figure}

With the improvement of detection efficiency we can demonstrate
violation of the criterion in Eq.~\ref{crit} directly using the
measured data. Figure~\ref{violation}a shows the difference
between the measured joint signal-idler photon-number distribution
$ f(n_S,n_I) $ and the calculated joint signal-idler photon-number
distribution given by a direct product of the signal and idler
marginal distributions. We can see in Fig.~\ref{violation} that
elements lying on a diagonal or near the diagonal are enhanced
whereas those lying far from the diagonal are suppressed. This is
a direct experimental manifestation of the fact that signal and
idler photons are generated in pairs. The correlation coefficient
$ C_p $ computed from the measured data shown in
Figure~\ref{violation}a is $C_p=0.051\pm0.005$. This value being
higher than that in our previous publication reflects an improved
detection efficiency. In Figure~\ref{violation}b we plot the
measured joint signal-idler photon-number distribution together
with a contour line around the point $n_s=8, n_i=9$ where the
criterion expressed in Eq.~(\ref{crit}) is violated by 2.8
standard deviations. A larger deviation can be obtained if a
longer measurement sequence is used and instabilities of the
measurement setup and fluctuations of the pump intensity are
lowered.

\section{A direct measurement of spatial correlations}

\begin{figure}
\begin{center}
\includegraphics[width=0.5\textwidth]{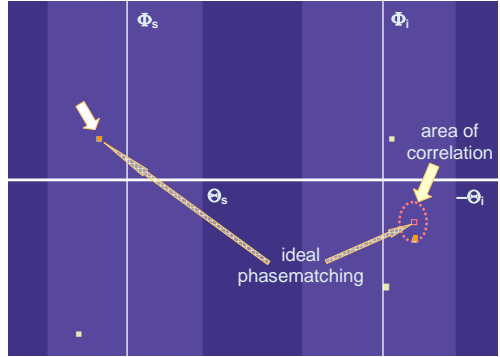}
\end{center}
\caption{\label{area}Definition of the area of correlation in the
detection plane of the iCCD camera.}
\end{figure}

As already pointed out in Introduction, the benefit of the
iCCD detector lies not only in its ability to be used as
a photon-number resolving detector but also in the fact that it
provides a spatial information about a detection event. In the
geometry of our experiment, the signal ($\Theta_s, \Phi_s$) and
idler ($\Theta_i, \Phi_i$) output angles approximately translate
to horizontal and vertical coordinates in the iCCD detection plane
($x_s,y_s$, $-x_i,y_i$); the minus sign in the idler strip comes
due to a horizontal inversion of the strip after reflection on the
mirror (see Figure~\ref{area}). Provided that the distance of the
camera from the crystal is large enough, the error introduced by
this transition from polar to rectangular coordinates is
negligible compared to the size of the detector macropixel
(several pixels of the iCCD can be grouped together into one macropixel
in the hardware of the camera to speed-up the image readout).

\begin{figure}
\begin{center}
\includegraphics[width=\textwidth]{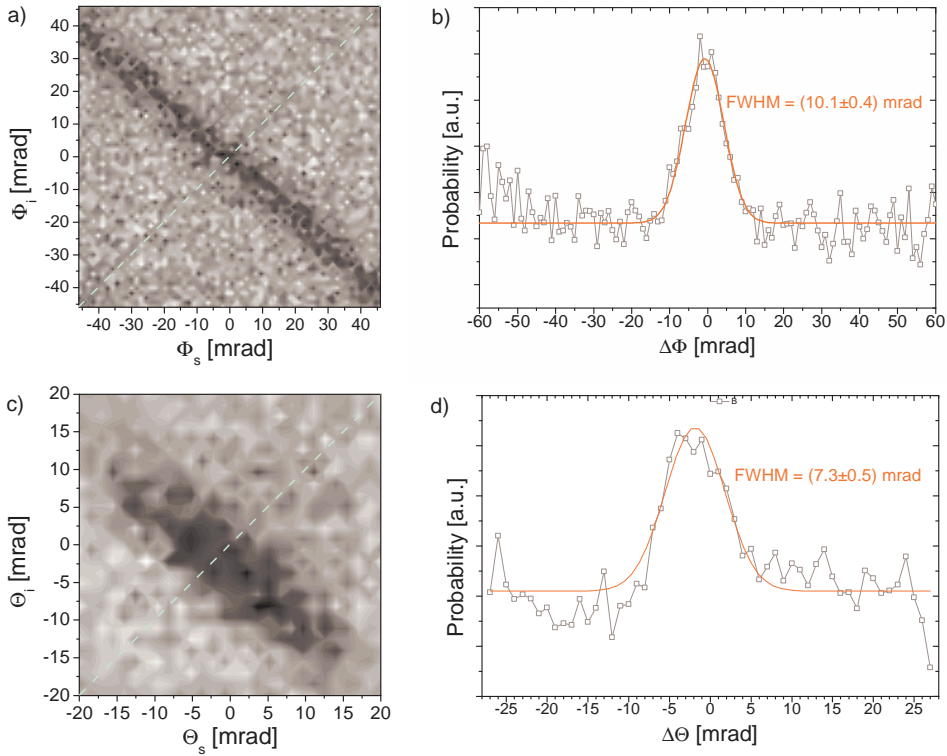}
\end{center}
\caption{\label{corrarea} Results of the measurement of the
dimensions of the area of correlation in the radial ($\Theta$) and
angular ($\Phi$) directions. Left pictures show signal-idler
plots of angular (a) and radial (c) positions of detection events.
The darkened diagonal is a manifestation of the correlation in the
positions of detection events. The plots on the right show
cross-section of the diagonal in a perpendicular direction and a
gaussian fit to the data. From the fits, the dimension of the
area of correlation in the angular (b) and radial (d) directions can
be obtained.}
\end{figure}

Since the total detection efficiency is quite low, we often detect
only one member of a photon pair (see Figure~\ref{area}). We do
not want to make a priori assumptions about the dimensions of the
area of correlation and so we take into account all possible
combinations of detection events in the signal and idler strips.
For example, if there is $n_s$ ($n_i$) detections in the signal
(idler) strip in a particular frame, we plot the total of $n_s
n_i$ points in Figs.~\ref{corrarea}a(c) at coordinates
[$\Phi_s(i),\Phi_i(j)$] ([$\Theta_s(i),\Theta_i(j)$]),
$i=1,..,n_s, j=1,..,n_i$, each with a weight of $1/(n_s n_i)$. The
plots are then accumulated along all the frames in one measurement
sequence. The resulting plots as shown in Figs.~\ref{corrarea}a,c
then clearly reveal correlations in the positions of detection
events in both coordinates. In both cases we can see a diagonal
coming from upper left to lower right corner of the plot. The
width of a diagonal corresponds to the dimension of the area of
correlation in the corresponding coordinate. We can see
cross-section of the diagonal in Figs.~\ref{corrarea}b,d. From
Gaussian fits to the measured data we can determine both radial
dimension ($7.3\pm0.5$ mrad, FWHM) and angular dimension
($10.1\pm0.4$ mrad, FWHM) of the area of correlation. Since
characteristics of the pump beam are not known with a sufficient
precision at present, a quantitative comparison of the measured
dimensions of the area of correlation with a theoretical model of
SPDC cannot be done.

We note that measurement of spatial correlations is much less
sensitive to instability of the pump intensity so that larger
sequences of camera frames can be processed within one
measurement.

\section{Conclusions}

% We have improved our measurement results concerning the joint
% signal-idler photon-number distribution using an iCCD camera compared
% to our previous publication \cite{Haderka05}.
By improving the detection efficiency of an iCCD camera with
respect to our previous results \cite{Haderka05} we have obtained
experimental data that directly violate the criterion of
classicality for a joint signal-idler photon-number distribution
introduced in \cite{Haderka05}. In addition, radial and angular
dimensions of the area of correlation have been experimentally
determined.

\section*{Acknowledgement}

The authors acknowledge support by the projects 6198959213 of the
Ministry of Education of the Czech Republic and 202/05/0498 of the
Grant Agency of the Czech Republic.

\end{document}